\documentclass[oneside]{article}
\usepackage[T1]{fontenc}
\usepackage{authblk}
\usepackage{float}
\usepackage{amsmath}
\usepackage{graphicx}
\usepackage{xfrac}
\usepackage[english]{babel}
\usepackage{lmodern}
\usepackage[T1]{fontenc}
\usepackage{xcolor}
\usepackage[latin9]{inputenc}
\usepackage{mathtools}
\usepackage{graphicx}
\usepackage{esint}
\usepackage[unicode=true,
 bookmarks=false,
breaklinks=false,pdfborder={0 0 1},backref=false,colorlinks=false]
 {hyperref}
\usepackage[a4paper, total={210mm,297mm},margin=2.0cm]{geometry}
\usepackage{setspace}
\usepackage{datetime}
\newdate{date}{2}{12}{2019}

\title{Multiparticle collision dynamics for fluid interfaces with near-contact interactions}

\author[1]{A. Montessori \thanks{Electronic address: \texttt{a.montessori@iac.cnr.it}; Corresponding author}}

\author[1]{M. Lauricella}

\author[1,2]{A. Tiribocchi}

\author[1,2]{F. Bonaccorso}

\author[1,2,3]{S. Succi}

\affil[1]{Istituto per le Applicazioni del Calcolo CNR, via dei Taurini 19, Rome, Italy}
\affil[2]{Center for Life Nano Science@La Sapienza, Istituto Italiano di Tecnologia, 00161 Roma, Italy}
\affil[3]{Institute for Applied Computational Science, John A. Paulson School of Engineering and Applied Sciences, Harvard University, Cambridge, USA}

\date{\displaydate{date}}

\begin{document}

\maketitle

\begin{abstract}
We present an extension of the multiparticle collision dynamics method for flows 
with complex interfaces, including supramolecular near-contact interactions
mimicking the effect of surfactants. The new method is demonstrated 
for the case of (i) short range repulsion of droplets in close contact, (ii) arrested phase separation 
and (iii) different pattern formation  during spinodal decomposition of binary mixtures.
\end{abstract}

\maketitle

\section{Introduction}

Multiphase and multicomponent flows are ubiquitous in nature and engineering applications and a 
thorough knowledge of the complex dynamics taking place  at the fluid interface level is fundamental 
to gain a deeper understanding of a variety of physical processes, such as combustion, 
material coatings and rain dynamics\cite{tryggvason2011direct,bates2005computational,schwarzkopf2011multiphase}, to name but a few.

Interface interactions, like coalescence and/or repulsion between droplets and bubbles, originate from 
the combined action of molecular attractive and repulsive forces, such as Van der Waals and electrostatic forces, steric 
interactions, hydration repulsion and hydrodynamic drag within thin films originating from the relative motion between fluid interfaces \cite{derjaguin1940,verwey1948,degennes2004}.

From a theoretical standpoint, to date, there is no systematic mathematical framework  capable 
of providing a full picture of the  interactions  at the fluid-fluid interface, which are known to be 
largely controlled by microscopic processes occurring between the molecules of two fluids in close contact \cite{montessori2019jfm,bergeron1999}.

A direct account of molecular forces in a milli-scale simulation requires to solve, simultaneously, six spatial decades ranging from tens/hundreds
micrometers, the typical size of droplets and bubbles in foams and emulsions, down to a few nanometres ,  the typical spatial scale of
near-contact forces \cite{montessori2019mesoscale,marmottantsoft,bergeron1999}. 

With the current state of affairs, such full scale simulations are still beyond reach even for the most powerful computers \cite{succi2019towards}, and such limitation
has motivated significant work in the direction of developing coarse grained models capable  of reproducing the 
correct interface physics while retaining the computational viability, thereby enabling the simulation 
of multiphase systems at experimental scales (mm to cm).

Among them,  mesoscale kinetic methods, such as the lattice Boltzmann method \cite{succi2018lattice,benzi1992,kruger2017lattice}, dissipative particle dynamics \cite{espanol1995statistical} 
and the multiparticle collision dynamics (MPCD) \cite{malevanets1999mesoscopic,kapral2008multiparticle,gompper2009multi,allahyarov2002mesoscopic}, have met with significant progress for the last three decades.

The MPCD technique, introduced by Malevanets and Kapral, consists of discrete streaming and collision steps 
and shares many features with both Bird's Direct Simulation
Monte Carlo (DSMC) \cite{bird1978monte} method and lattice gas models \cite{frisch1986lattice,succi1988high}.
Indeed, collisions occur at fixed discrete time intervals, and although space is discretised into cells,
both spatial coordinates and  velocities of the particles are grid-free.

The MPCD has been also generalised to model binary mixtures \cite{hashimoto2000immiscible,inoue2004mesoscopic} by assigning (i) a color charge,
 $c_i= \pm 1 $, to two different species of particles and (ii) by defining a rotation angle in the collision process, chosen 
 such that the color weighted momentum in a cell is rotated to point in the direction of the gradient of the color field.
 
A similar approach has been previously proposed within the lattice gas automata framework by Rothman et al. \cite{rothman1988immiscible}.
 
In this paper, we extend the MPCD-based approach for binary mixture with an additional rotation term, which 
is aimed at representing  the repulsive action of  near-contact forces operating at the fluid interface level. 
Such forces are due to the presence of amphiphilic groups and colloids at the fluid interface, as they
are known to delay or even arrest the interface coarsening in binary systems, as well as the coalescence of droplets 
in dense emulsions.

The proposed approach is shown to be capable of i) preventing the coalescence of fluid interfaces in close contact, 
ii) simulating the arrested phase separation during spinodal decomposition and iii) reproducing different structures 
of arrested phase binary mixture.

Taken all together, this is expected to enable the application of the MPCD method to a much broader
class of problems than previously amenable.

\section{Method}

In the following we discuss the extension of the MPCD technique to the case of simulations of the
canonical ensemble, using the Andersen thermostat. 

\subsection{Multiparticle Collision Dynamics with Andersen Thermostat}

In the multiparticle collision dynamics, the fluid is modelled by means point-like particles of mass $m$. 
The number of particles is typically pretty large ranging between $10^2 \div 10^3$ per linear dimensions.

Each simulation time consists of the repeated application of two basic dynamic processes: streaming and local relaxation.

During the streaming step, particle positions are updated via a forward Euler step:

\begin{equation}
\vec{r}_k(t + \Delta t)=\vec{r}_k(t) +\vec{v}_k(t)\Delta t + \frac{1}{2m} \vec{f}_{ext}\Delta t^2
\end{equation}

where $\vec{r}_k$ is the position of the $k-th$ particle, $\Delta t$ is the MPCD time-step, $m$ the mass of 
a particle and $\vec{f}_{ext}$ is an external forcing.

As per the relaxation process, several approaches are available \cite{gompper2009multi}, but in this work we chose 
to perform the MPCD simulation directly in the canonical ensemble by employing an Andersen Thermostat (AT) \cite{weinan2008andersen,gompper2009multi}.

In the MPCD-AT, new relative velocities are generated during each computational step in each cell as follows:

\begin{equation}
\vec{v}_k(t+\Delta t)= \vec{v}_{cm}(t) + \delta \vec{v}_k^{ran}=\vec{v}_{cm}(t) + \vec{v}_k^{ran} - \frac{1}{N_c}\sum_{j \in cell} \vec{v}_j^{ran}
\label{MPCD-AT}
\end{equation}

where $\vec{v}_k^{ran}$ are random numbers sampled from a Gaussian distribution with variance $\sqrt{k_BT/m}$ 
and $N_c$ is the actual number of particles in the collision cell.
In the above equation, the sum runs over all particles in a given cell.

Moreover, an analytical expression for the kinematic viscosity of an MPCD-AT fluid can be derived, which reads \cite{kikuchi2003transport,gompper2009multi}:
\begin{equation}
\nu= k_BT\Delta t \left[ \frac{N}{N-1 + e^{-N}} - \frac{1}{2} \right] + \frac{\Delta x^2}{12 \Delta t} \left[ \frac{N-1 +e^{-N}}{N}  \right]
\end{equation}
where $N$ is the average particle density in a cell and $\Delta x$ is the length of the side of a square MPCD cell.

The choice of the Andersen thermostat is motivated by the fact that, although computationally more expensive 
than the standard stochastic rotation dynamics approach, it features shorter relaxation times, thus reducing the 
number of time steps required for transport coefficients to reach their asymptotic values \cite{kapral2008multiparticle}. 

In the next section, we detail the methodology employed to trigger a phase-separation in an MPCD -AT fluid.

\subsection{Multicomponent model with near-contact interactions}

The multiparticle collision dynamics can be extended to multicomponent fluid systems, by making 
use of 'colored' particles, which interact via simple rules which code for attraction between particles 
of the same color (i.e. red-red) and repulsion between particles of different color (i.e. red-blue or A-B).  

This idea was proposed by Rothman et al. \cite{rothman1988immiscible} for FHP lattice gas for the simulation
of phase separation .

\textcolor{black}{It is worth noting that this is not the only possible choice to simulate binary fluids within the MPCD framework(see \cite{ihle2006consistent} for example).\\
For instance, recently, Echeverria et al. \cite{echeverria2017mesoscopic}, proposed a multiparticle collision
model with a repulsion rule which aimed at  investigating phase separation processes in  binary fluids in both free and crowded
environments. }

The algorithm can be viewed as an extension of the collision process.

First, we proceed by defining a set of vectors between a cell center ($\vec{x}$) and its neighbor 
cell centers ($\vec{x}_i$, $i=1,8$ in two dimensions):

\begin{equation}
\vec{d}_i=\vec{x}_i -\vec{x}
\end{equation} 

where $i$ spans the directions of a $4^{th}$- order isotropic lattice in two dimensions (i.e. a nine speed stencil \cite{thampi2013isotropic}).

One then defines the color weight of the $n-th$  particle in a cell as:

\begin{equation}
p_n=\begin{cases}
               +1 \; \textrm{if the particle is red} \\
               -1 \; \textrm{if the particle is blue}
            \end{cases}
\end{equation}

Upon summing the color weights over the particles within a given cell,  one determines whether the given
cell belongs to the red or blue fluid. 

The cell type can be then defined by computing the following quantity:

\begin{equation}
\Delta(\vec{x})=\sum_{n\in \vec{x}} p_n
\end{equation}

the index $n$ running through all the particles within the  cell centered in $\vec{x}$.

Next, we introduce two additional vectors, namely the color flux $\vec{q}$ and the color field $\vec{f}$.

The color flux, $\vec{q}=\sum_n p_n(\vec{v}_n - \vec{v}_{cm})$, is a color-weighted momentum of the cell, while 
 $\vec{f}= \kappa_{\sigma} \sum_{l} w_l \Delta(\vec{x} + \vec{c}_l)\vec{c}_l$  is the color gradient between neighbor 
cells,  $\kappa_{\sigma}$ being the parameter controlling the magnitude of the interaction force, i.e. to set the surface tension of the model.
Finally, $w_l$ is a set of weights which guarantees the isotropy of the gradient operator \cite{thampi2013isotropic} and $\vec{c}_l$ are, in two dimensions, 
the nine lattice directions, supporting the gradient calculation.
 
We then define a rotation matrix $\Omega$, designed so as to guarantee the overlap 
between the color flux and the color gradient after the collision process:
\begin{equation}
\frac{\vec{f}}{|\vec{f}|}=\Omega \frac{\vec{q}}{|\vec{q}|}
\end{equation} 

In this way, particles of the same color are forced to gather together and to move towards nodes where particles 
of the same kind are in majority, which is tantamount to triggering phase-separation in the binary MPCD model. 

The model presented so far only allows  the simulation of fluid mixtures in which interfaces of the same fluid 
naturally coalesce with each other, under the drive of energy minimisation (least surface/volume ratio). 

In other words, the current model does not incorporate any short range disjoining force able to (i) suppress 
the coalescence between fluid interfaces in close contact and (ii) to sustain the thin film which forms between them. This occurs, for example, in surfactant of polymer solutions, in which the molecular interactions between 
amphiphilic groups give rise to repulsive forces between the interfaces and impede or completely arrest drainage within thin films \cite{derjaguin1940}. 

The aim is then is to modify the binary MPCD model to account for short range repulsion between fluid interfaces by introducing a local repulsive force aimed at providing a mesoscale (coarse grained) representation of the near contact forces (i.e. van der Waals, electrostatic, steric and hydration repulsion) acting between fluid interfaces in close proximity.

The additional collision term is applied only to those cells belonging to a thin film between two fluid interfaces (i.e. a blue cell between red majority cells). 
The thin film condition is very easy to assess, via the use of the $\Delta(\vec{x})$ function which can be employed to characterize 
whether a cell is A majority or B majority, being A and B the two components of the system. 
By computing $\Delta$ for each cell, one is able to draw a map of the cell types within the MPCD system.
 
Thus, if $\Delta(\vec{x})<0$ (B majority) and $\sum_l \Delta(\vec{x} + 3\vec{c}_l)=\Delta(\vec{\hat{x}})\geq 6$ 
(i.e., at least six of the surrounding cells at a distance $3 \Delta x$ , being $\Delta x$ the width of an 
MPCD cell, are A majority)  the current cell, $\vec{x},$ is marked as a thin film cell. 

We then compute the repulsive forcing term as :

\begin{equation}
\vec{f}_{rep}= \sum_i \kappa_{rep} w_i \Delta(\vec{\hat{x}})
\end{equation}

where $\kappa_{rep}$ measures the strength of the short range repulsive interaction .

The local repulsive force is then applied in the collision step via an (extended) 
rotation matrix, which is reported below for the sake of clarity:

\begin{equation}
\Omega= \begin{pmatrix}
(q_x f_x - q_x f_{rep,x}) + (q_y f_y - q_y f_{rep,y}) & (-q_x f_y + q_x f_{rep,x}) + (q_y f_x - q_y f_{rep,x}) \\ 
(q_x f_y - q_x f_{rep,y}) - (q_y f_x - q_y f_{rep,x}) & (q_y f_y - q_y f_{rep,y}) + (q_y f_y - q_y f_{rep,y})
\end{pmatrix}
\end{equation}

\section{Results}

In this section we test the extended MPCD model on three applications namely, the brownian motion of a droplet in a fluctuating fluid,
the short range repulsion between droplets and the arrested phase separation in spinodal binary mixtures.

\subsection{Brownian motion of a droplet in a fluid at rest}

As a first test, we simulated  the Brownian motion of a  droplet immersed in a fluctuating bulk fluid. 
The simulation was performed on a $136 \times 136$ fully periodic grid.

The simulation setup consists of droplet of radius $15 \Delta x$ in MPCD units,  surrounded by a bulk fluid of the same density and viscosity. 
The positions and velocity of the dispersed fluid (i.e. the droplet) are randomly initialized in a region of space  within the radius of the droplet,
while the bulk particles are randomly positioned outside it.

The thermal energy of the system is set to $k_BT=0.1$, the average cell density is $N=10$ and 
the surface tension of the model, obtained via the Laplace test, is $\sigma\sim 0.1$.

\begin{figure}
\begin{center}
\includegraphics[scale=0.6]{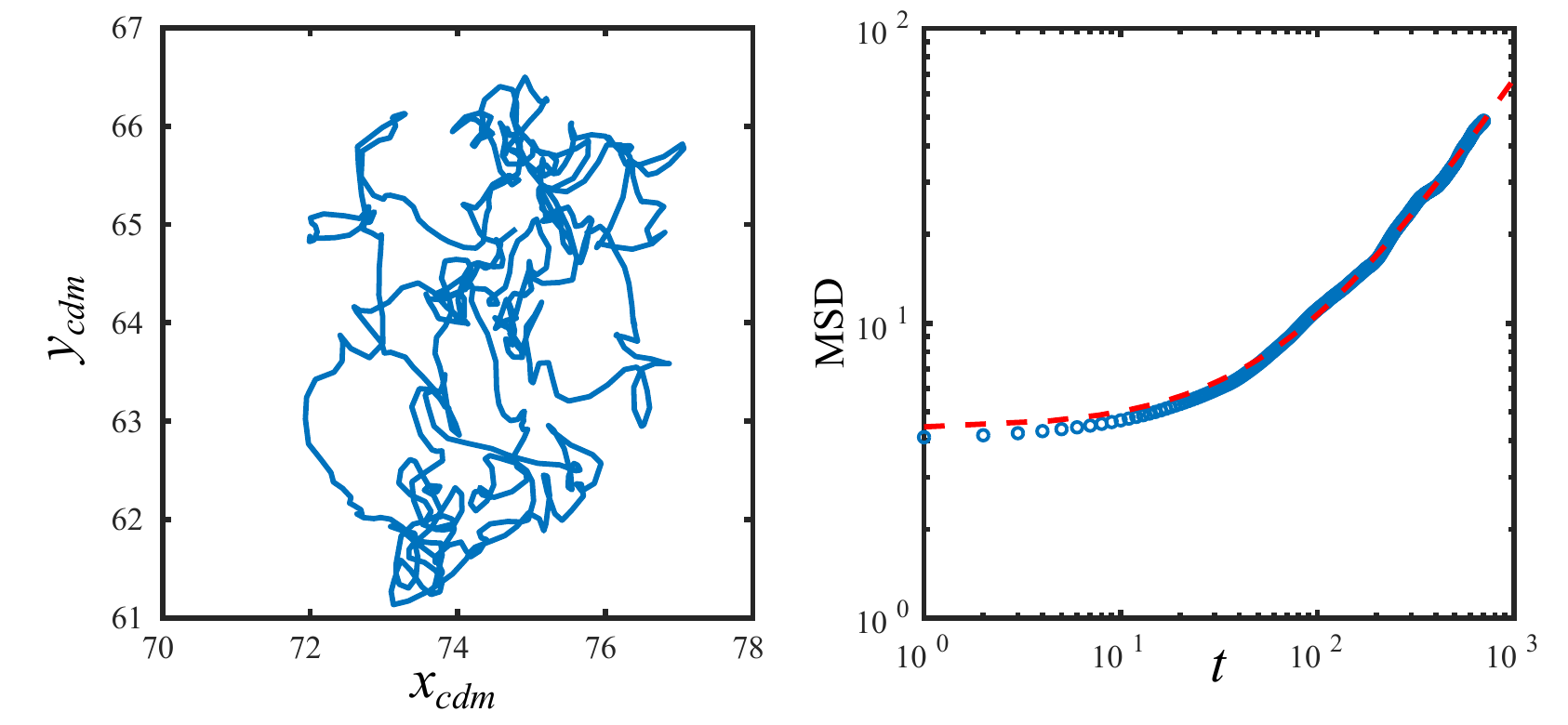}
\caption{\label{brownian} (Left panel) Brownian path of the droplet center of mass driven by thermal fluctuations in the bulk and (right panel) Mean Squared Displacement of the center of mass position vs time. numerical simulation (Symbols), dashed line (linear fit). The diffusivity of the droplet in the fluctuating bulk fluid, evaluated as  the angular coefficient of the interpolating linear function, is $\sim 0.0315$ (in MPCD units).  }
\end{center}
\end{figure}

Due to the thermal fluctuations in the bulk fluid, the droplet performs a Brownian path around 
the center of the domain, as clearly shown in the left panel of figure \ref{brownian}.

To confirm that the droplet follows a random Brownian motion, we computed the mean square displacement, 
defined as $\frac{1}{N_t} \sum_t^{N_t} |\vec{x}_{cdm}(t)-\vec{x}_{cdm}(0)|^2$, $N_t$ being the number
of simulation steps and $\vec{x}_{cdm}$  the center of mass position vector.

As evidenced by the log-log plot reported in the right panel of figure \ref{brownian}, the mean squared displacement is a linear function of time, this confirming the Brownian nature of the droplet motion in the thermal bath.

From the same plot, it is also possible to compute the diffusivity of the droplet within the bulk fluid, defined as the slope
of the linear fit, which gives $\sim 0.0315$ in simulation units.

We also computed the probability distribution function (PDF) for the velocity of the droplet by recording  the velocity of the center 
of mass of the droplet (not shown in figure).

As expected, the histograms agree with the Maxwell-Boltzmann distribution at the given temperature, 
further confirming the random diffusive path followed by droplet under the effect of the thermal fluctuations of the background fluid.

\subsection{Short range repulsion between droplets in close contact}

In this subsection, we show that the inclusion of the short range repulsion in the multicomponent MPCD 
frustrates the coalescence between fluid interfaces in near contact.

The simulation setup consists of a fully periodic box, in which two  droplets of radius $R$, surrounded by a bulk fluid 
with the same density and viscosity of the dispersed phase, are placed at a close distance.

The surface tension of the mixture $\sigma \sim 0.16$ was obtained by setting $k_BT=0.1$ and we use an average cell 
density $N=60$. The strength of the repulsive force was set to $\kappa_{rep}=-1.5$.
\begin{figure}
\begin{center}
\includegraphics[scale=0.7]{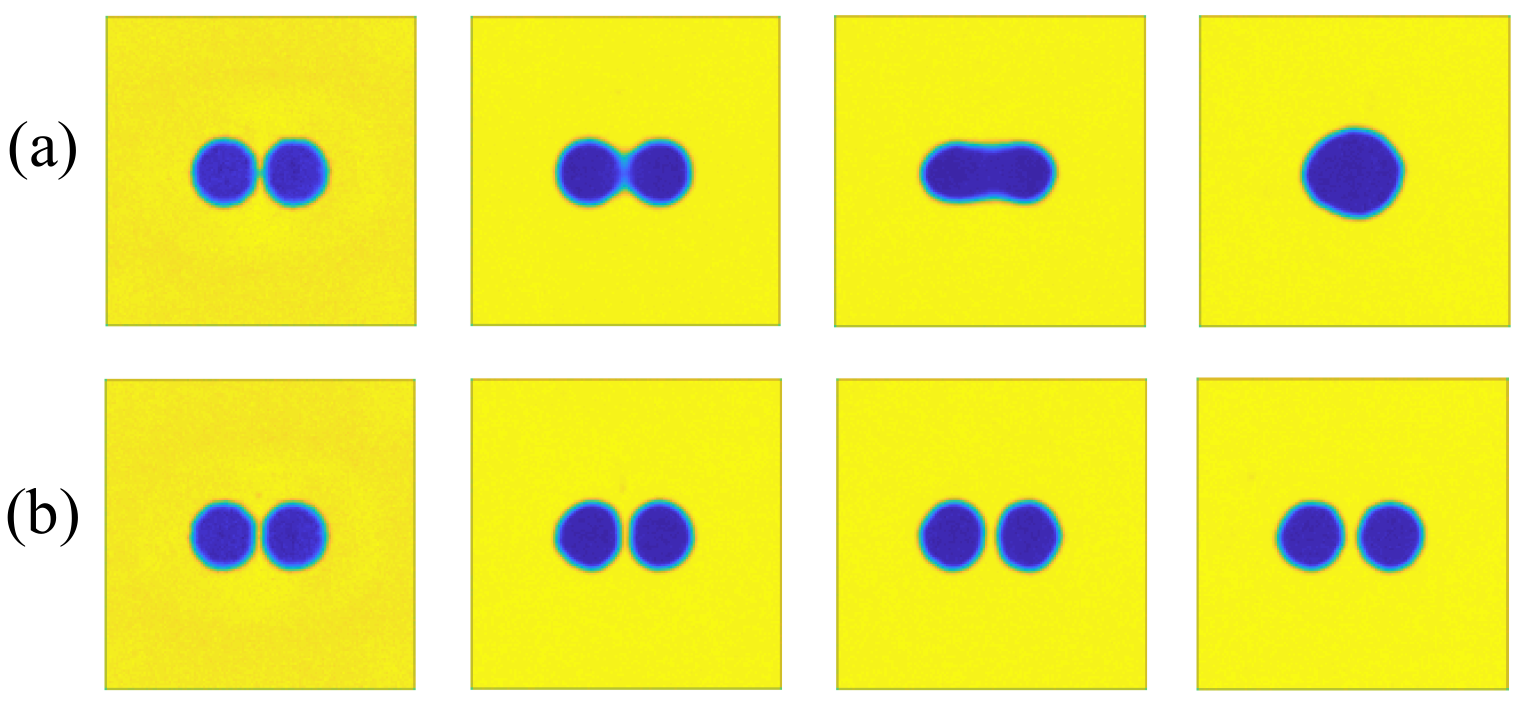}
\caption{\label{repulsion} Sequences of (a) coalescence and (b) repulsion between two droplets in a surrounding fluid at rest.}
\end{center}
\end{figure}

As shown in figure \ref{repulsion} (a), in absence of repulsive near contact forces, the two droplets spontaneously coalesce, forming 
a bridge which grows in time until the formation of a single droplet. 
The coalescence occurs spontaneously, since the system evolves under  the constraint
of minimising its interfacial energy.

The same simulation was then performed with the repulsive force on, which is intended to simulate 
the repulsion due to the presence of amphiphiles at the fluid-fluid interface.

As evidenced in figure \ref{repulsion}(b), the addition of the repulsive force within the thin film is 
sufficient to induce a short range repulsion between the droplets with a subsequent frustration of the coalescence between them. 

\begin{figure}
\begin{center}
\includegraphics[scale=0.5]{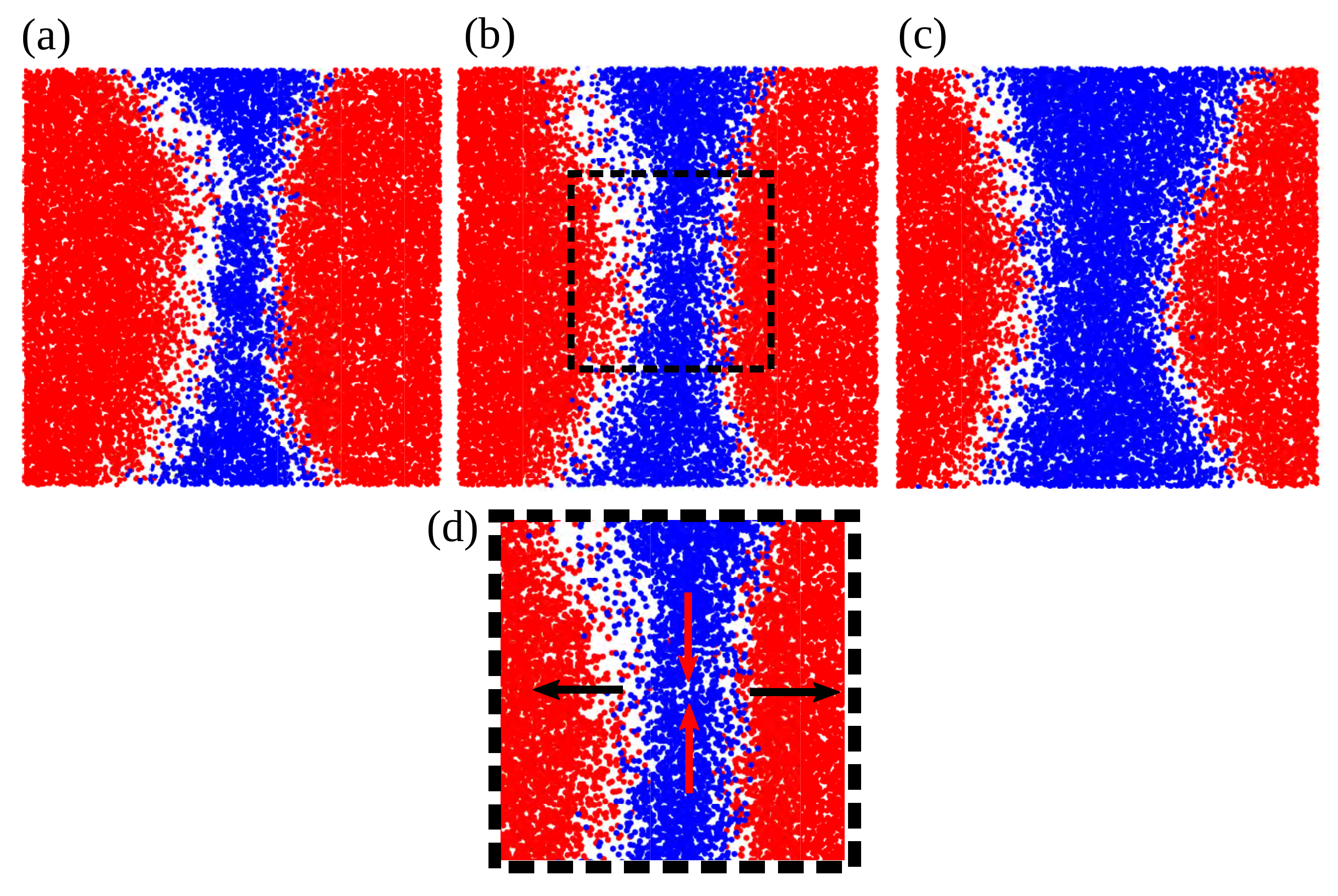}
\caption{\label{part_rep} (a-c) Scatter plots of the repulsive process between two droplets in close contact. (d) Zoom of the thin film. Black arrows denotes the repulsive forces between particles of the dispersed phase while the black arrows indicate the recirculation of the particles in the thin film. The particles of the continuous phase flow from the periphery rather than escaping towards the bulk, stabilizing the thin film and determining the repulsive force between the droplets.  }
\end{center}
\end{figure}

In other words, the effect of the forcing term is to stabilise the thin film between fluid interfaces in close contact by inducing 
a local positive (i.e. attractive) interaction between particles of different species.

In this way, the bulk fluid particles are forced to recirculate within the thin film, thus preventing the film rupture which, ultimately, causes 
the coalescence between the droplets. 

This process is evidenced in figure \ref{part_rep}, which shows the scatterplots of the continuous (bulk) and dispersed (droplets) phase.

In the zoom, the arrows sketched the two processes which take place within the thin film and at the interface of the two droplets in close contact. 
The repulsive force determines a counter-flow of the continuous phase particles from the periphery of the thin film towards its bulk. 
This counter-motion prevents the drainage of the thin film and determines a repulsion between the droplet interfaces, which
consequently move away from each other.

Since the particles of the continuous phase move from the periphery to the bulk of the thin film, it is possible 
to detect a local  increase of the cell density, which acts as an effective disjoining pressure within the fluid thin film.

When the repulsive force is switched off, the interaction between particles of different colours (A-B) is repulsive, thus the fluid 
particles included in the thin film are naturally depleted from it, thus promoting the formation of a fluid bridge between the interfaces in close contact. 

\begin{figure}
\begin{center}
\includegraphics[scale=1.]{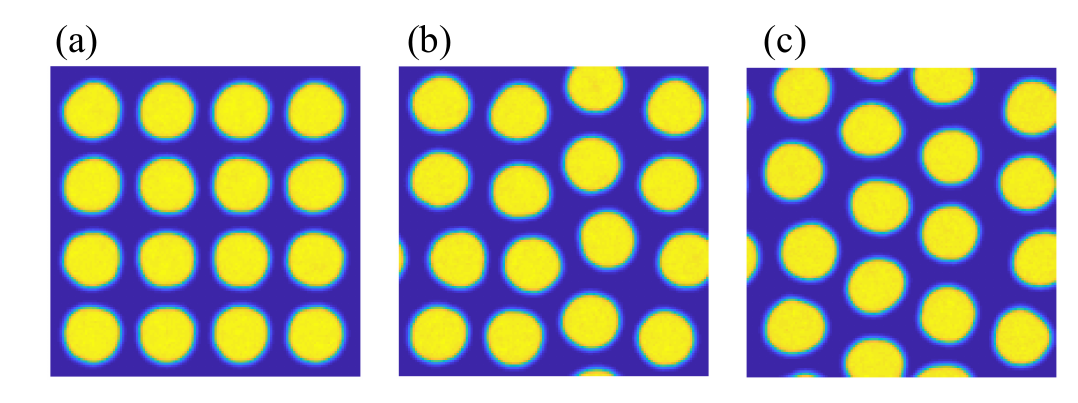}
\caption{\label{emu} (a-c) time sequence of the droplets diffusion of a dense emulsion in a periodic box.   }
\end{center}
\end{figure}

As a further test, we performed a simulation of a dense emulsion formed by a cluster of $16$ immiscible droplets, suspended 
in a fluctuating background fluid (figure \ref{emu}). 
Also in this case, the domain is a fully periodic square box of side $152$.

The simulation starts with the the droplets  placed on a regular array,  each center being surrounded by 
other four at a distance of one droplet diameter.
The droplets are then left free to move within the fluctuating bulk fluid.

As expected, the droplets start to move around their initial position, performing a random motion driven 
by the thermal agitation of the surrounding fluid, occasionally interacting with each other without coalescing.

As a result, the repulsive force allows the stabilisation of the emulsion by suppressing the attractive interaction  
between the particles of the same fluid component.

It is interesting to note that the present approach permits to model comparatively large emulsions, with hundreds of droplets or more, 
by employing  just two species of particles, at variance with other models which track the 
evolution of $N$ immiscible droplets in a fluctuating fluid by introducing $N+1$ different species and, consequently, $N$ different collision rules \cite{hashimoto2000immiscible}.
From this standpoint, the proposed model provides an efficient alternative to model multi-droplet interactions with built-in thermal fluctuations.  

In the next subsection, we show that the inclusion of near contact interactions allows to reproduce an
 arrested phase separation in spinodal decomposition of binary fluid mixtures.
 Incidentally,  these simulations exhibit a rich variety of different patterns and configurations, typical
 of  multi-component systems in the presence of amphiphilic groups or colloidal particles (i.e. bijels \cite{tiribocchi2019curvature,harting2005large}).

\subsection{Spinodal decomposition with arrested phase separation}

In this section, we discuss the effects of the near-contact repulsion in a spinodal decomposition of a binary mixture.
\textcolor{black}{Phase separation in spinodal decomposition occurs whenever a
mixture  enters into the unstable region of the phase diagram . The boundaries of the unstable
Region (the binodal region), can be obtained by performing a
Maxwell-Construction, while the region within the binodal curve is called the
spinodal.
The binodal and spinodal curves meet at the critical point.\\
When a fluid mixture transits into the spinodal region of the phase diagram it undergoes to a spinodal decomposition.\\
In this work, we performed spinodal decomposition of both critical  and off-critical quenches with and 
without the inclusion of the near contact forcing term.\\
As shown below the combined effect of the volume fraction, the temperature and the presence of the near contact force gives rise to a number of different patterns 
as the final outcome of the spinodal process.
}
As before, the simulation setup  consists of a fully periodic box
in which both the particle positions and velocity of two components $A$ and $B$ are randomly initialised.
The system is then left free to evolve without any external body force.
Also  in this case, the surface tension of the mixture was fixed at a constant
value, $\sigma = 0.08$, corresponding to $k_BT=0.05$, the average cell density is $N=60$ and, when the repulsive force is switched on, $\kappa_{rep}=-1.5$.

\begin{figure}
\begin{center}
\includegraphics[scale=0.7]{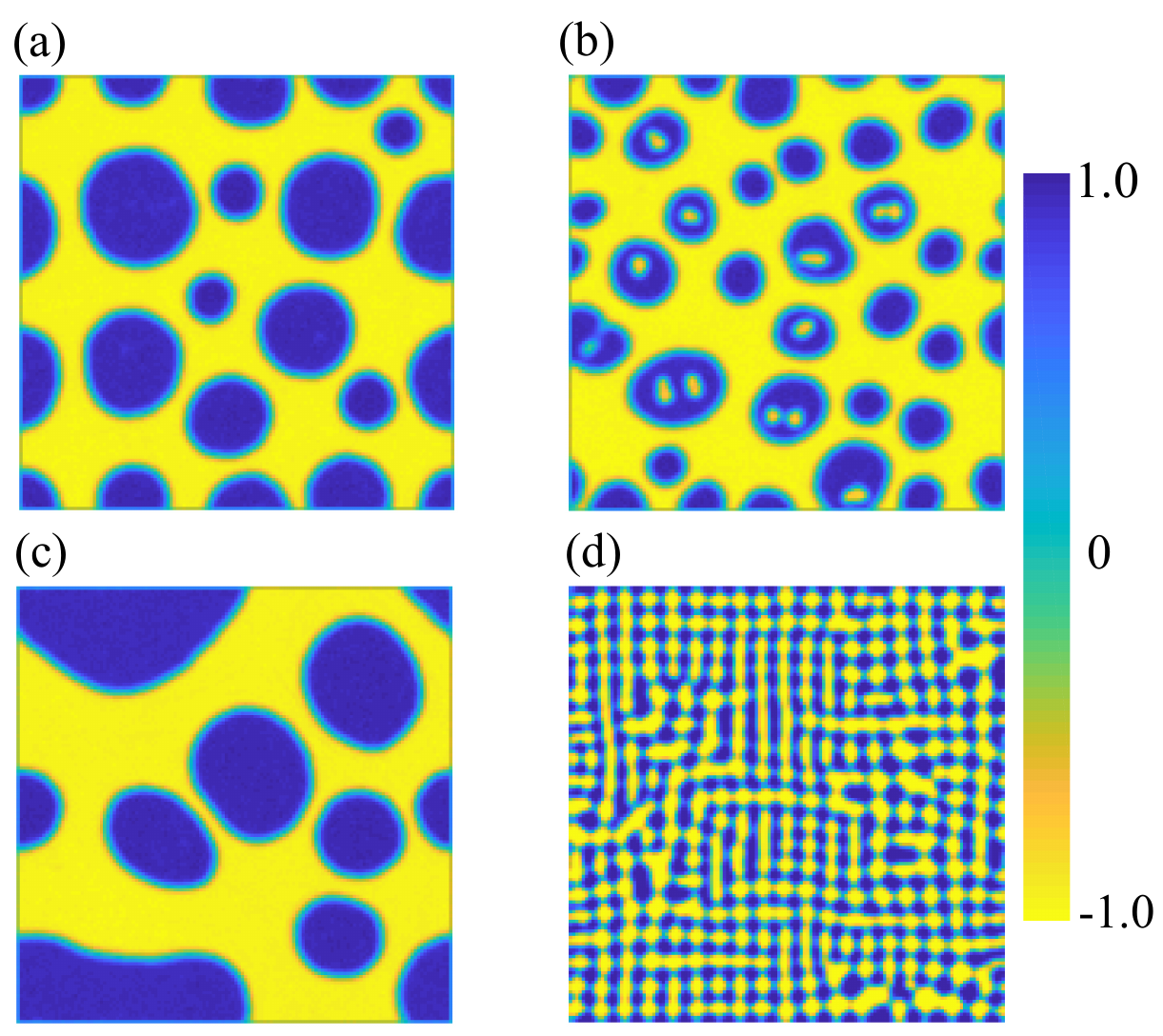}
\caption{\label{spinodal} Spinodal decomposition without ((a) and (c)) and with the near contact forces on ((b) and (d)) for two different values of volume fractions ($N_A/N_{tot}$). \textcolor{black}{(a) and (b) off-critical quenches,(c) and (d) critical quenches}. The repulsive short range force promote the separation of the two fluids by arresting the coarsening between the phases.}
\end{center}
\end{figure}

In figure \ref{spinodal}, we report four snapshots of the fluid system after $3\cdot 10^4$ simulation steps for four different cases.

In particular, on the left side ((a) and (c)), the fluid system is evolved without incorporating the near contact force between fluid-fluid interfaces 
and the system has been initialised with two different values of volume fractions,$\phi=N_A/N_{tot} \sim 0.3$ \textcolor{black}{(off-critical quench)} $\phi=N_A/N_{tot} \sim 0.5$ \textcolor{black}{(critical quench)} respectively, being $N_R$ the number of particle of the droplet phase within the system and $N_{tot}$ the total number of particles . 
On the right hand side, we report the same simulation, but with the short range repulsive force on.

As one can see, the effect of the interface repulsion is quite significant, as evidenced by  the phase fields reported in fig. \ref{spinodal}(a) and \ref{spinodal}(b). 

In the first case (no repulsion), we observe a standard spinodal separation, in which the binary system evolves 
in such a way that the  fluis molecules immediately start to cluster together into microscopic A-rich and B-rich clusters throughout the liquid.
 
These clusters then rapidly grow and coalesce, forming bigger droplets, which eventually coalesce with each other until the formation  
of two separated  macroscopic B-rich and A-rich phases. 
This long-term coarsening can be observed by running the simulation longer.  
As before, the system naturally evolves towards the minimum energy configuration, which is represented 
by the final coarsening of the binary system.

On the other hand, the right upper panel clearly shows an arrested phase separation between the A and B phases. 
Indeed, at the same simulation step, the binary mixture presents smaller droplets, which cannot coalesce 
due to the presence of the short range repulsion between fluid interfaces. 
More interestingly, the phase field reveals the formation of a double emulsion (A-B-A), which  emerges spontaneously during the coarsening process. 
It is also interesting to note that, in this case, the spinodal decomposition is arrested, i.e. no more coarsening occurs between the interfaces 
after this point in time.
Indeed, the droplets keep moving within the periodic box due to the Brownian motion induced by the thermal fluctuations, but the 
coalescence between them is frustrated by the short-range repulsion.  

The arrested phase separation is even more pronounced in the second case ($\phi=0.5$), as shown in fig. \ref{spinodal}(d).
In this case, the result of the spinodal decomposition is a system composed by small droplets which align 
with each other and eventually coalesce to form stripes of A and B phases. In this case, the system is more dynamic, presenting 
some interesting out-of-equilibrium patterns in which the droplets split and coalesce with each other dynamically as time unfolds.

\begin{figure}
\begin{center}
\includegraphics[scale=0.7]{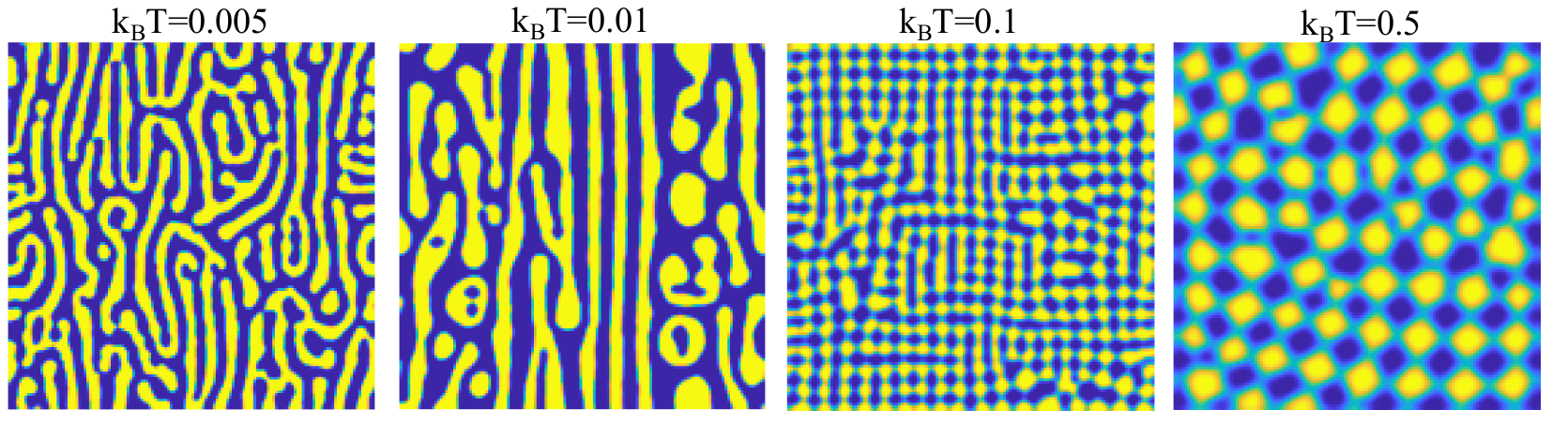}
\caption{\label{lamellar} Transition from lamellar to emulsion ordering \textcolor{black}{in a spinodal decomposition of a critical quench} for increasing values of the  temperature.}
\end{center}
\end{figure}

We then investigate the effects of the temperature on the binary system during the decomposition.
To this aim, we run three simulations at a fixed volume fraction $\phi=0.5$ and change the thermal energy of 
the system in the range $k_BT=0.01 \div 0.5$. All simulations are performed with the near-contact force on.

Figure \ref{lamellar} shows that the system temperature has a major impact on the spinodal separation and 
the arrested coarsening. 
Indeed, at low temperature, the system presents phase fields patterns typical of lamellar
phases, showing interesting labyrinth structures often met in metal alloys or ferrofluids \cite{jansen2011bijels,xu2006morphologies,gonnella1997spinodal,martins1998phase,schwarz2002bicontinuous,gompper1994phase}.

As the temperature increases, the system gradually looses the lamellar structure in favour of a droplet-droplet ordering, typical of emulsion-like systems.  

It is interesting to note that the rich variety of patterns described above, emerge  spontaneously from the simulations, 
by simply varying the temperature of the binary fluid system. 

A quantitative analysis of these phenomena, which involve the competition between surface tension, repulsive forces and thermal fluctuations, will make the subject of future investigation.

The proposed approach may be employed to shed new lights on the equilibrium and non-equilibrium 
behavior of complex binary systems with  colloidal particles or amphiphilic groups, including 
thermal effects typically not accounted for in standard Cahn-Hilliard models \cite{liu2015dynamic,GouyetPRE1995,wittkowski2014scalar} or in other hydrodynamic 
mesoscale approaches for complex flows \cite{swift1996lattice,swift1995lattice}.

\section{Conclusions}

Summarizing,  we have presented an extension of the multiparticle collision dynamics approach 
to model the effect of near contact interaction  forces between fluid interfaces in close contact. 

The model has been shown to effectively frustrate the coalescence between droplets suspended in a fluctuating background fluid.
The net effect of the near contact repulsive local force is to avoid the outward motion of the particles contained in thin films between neighbor fluid interfaces.
 
This, in turn, codes for the effect of an effective positive disjoining force, which is needed to avoid the coalescence between droplets 
and sustain the thin film between them.
Moreover, it has been shown that the incorporation of the short range repulsive force leads to the 
emergence of rich variety of patterns in spinodal binary mixture \cite{gonnella1997spinodal}.
Similar patterns in binary mixtures can be simulate by means of other mesoscale models, such as the lattice Boltzmann 
model with free energy for binary mixture. However, such models do not generally incorporate the effects of  
thermal fluctuations, or they do it to a very limited extent.
In this respect, the proposed model may open new computational routes for the investigation of complex 
emulsions and more exotic mesoscale states of matter, which result from a tight balance between
thermal effects and near-contact forces.

\section*{Acknowledgements}
A. M., M. L., A. T. and S. S. acknowledge funding from the European Research Council under the European Union's Horizon 2020 Framework
Programme (No. FP/2014-2020) ERC Grant Agreement No.739964 (COPMAT).


\end{document}